\documentstyle[12pt,aaspp4,rotating]{article}
\input{psfig.sty}
\def\lax    {${_<\atop^{\sim}}$}

\def\etal   {{\it et al.}~}

\begin{document}
\title{Multiple Velocity Components in the CIV Absorption Line of NGC5548$^1$}
\author{Smita Mathur$^2$, Martin Elvis, Belinda Wilkes}
\affil{Harvard-Smithsonian Center for Astrophysics,
60 Garden St., Cambridge, MA 02138}
\footnotetext[1]{Based on observations with the NASA/ESA {\it Hubble
Space Telescope},
obtained at the Space Telescope Science Institute, which is operated
by the Association of Universities for Research in Astronomy, Inc.,
under NASA contract NAS5-26555.}
\footnotetext[2]{smita@cfa.harvard.edu}

\setcounter{footnote}{1}

\received{}
\accepted{}
\lefthead{Mathur \etal~}
\righthead{NGC5548: GHRS}

\section*{Abstract}

We have observed the much studied Seyfert 1 galaxy NGC5548 with
the GHRS spectrometer on HST. Our 14ksec observation covers the
CIV emission line at a resolution of $>$20,000. Our purpose
was to study the absorption line found at lower resolution by IUE
and HST/FOS. We found
that the CIV absorption line resolves into 6 separate doublets with EWs of
0.07\AA--0.38\AA. The absorption lines have blueshifts relative to the
 systemic velocity of the galaxy
 of 380 -- 1250~km~s$^{-1}$, except for one
which has a redshift of 250~km~s$^{-1}$, suggesting both inflow
and outflow. The inflowing component may be related to the accretion
flow into the nuclear black hole. All the doublet lines are resolved
by the GHRS. Three doublets
are narrow with FWHM\lax100~km~s$^{-1}$, and three are broad, FWHM
$\sim$160--290~km~s$^{-1}$  . We find
evidence of partial covering by the narrow absorption lines. Either
(but not both) of the two strongest broad doublets could be
from the same material that produces the X-ray ionized
absorber seen in soft X-rays. The remaining five systems must be at
least 10 times
less ionized (and so of lower total column density) to remain
consistent with the X-ray spectra.


\newpage
\section{Introduction}

The bright, variable, Seyfert~1 galaxy NGC~5548 has been extensively
studied at many wavelengths.  It has been a target of reverberation
mapping experiments in the optical and UV (Peterson et al.~1992,
Clavel et al.~1991, Korista et al.~1995).  These have led to the
accurate determination of the physical size of the BELR.  The UV
spectrum also shows absorption lines [Shull \& Sachs 1993, Mathur,
Elvis \& Wilkes 1995 (MEW95)] and monitoring observations place upper
limits on the response time of the UV absorbers to changes in the UV
continuum.  The X-ray spectrum shows an ionized absorber (Nandra et
al.~1991, Reynolds 1997, George \etal~1998). NGC5548 is radio-quiet,
like most Seyfert galaxies, but it is not radio silent. Large radio
lobes are observed in the plane of sky indicative of an edge-on
nuclear geometry.

Recently, based on ASCA and HST FOS data, MEW95 showed that the
ionized X-ray and UV absorption in
NGC5548 is likely to originate in the same material. This unification
allowed us to combine the X-ray and UV constraints
to determine the physical conditions of the absorber.
In the FOS data, the measured line widths for the CIV doublet are
$\sim$350 km~s$^{-1}$ (FWHM), broader that the nominal spectral
resolution for this observation ($\sim$250 km s$^{-1}$) and thus
possibly resolved. This would imply that the absorber is dispersed in
velocity space.
High resolution observations would allow accurate determination of the
velocity spread parameter and resolve multiple components if any. This
in turn would allow better estimates of column density. We observed
the CIV line in NGC5548 with GHRS on HST with this aim in mind.

\section{Observations and Analysis}

We observed NGC5548 with GHRS on HST on 24 August 1996. The
exposure was centered around the CIV emission line. A total exposure
time of 14,035 seconds was obtained using G160M in ACCUM mode covering
the wavelength range 1554\AA--1590\AA. ~The
Large Science
Aperture ($1.74^{\prime\prime}\times1.74^{\prime\prime}$) was used,
providing a spectral resolution of $\approx20000$.
We also observed with the lower resolution G140L grating for 2176 seconds
to measure the complete line profile and part of the continuum on both
sides of the line (1417\AA\--1713\AA).~

Data reduction was performed using {\bf STSDAS} in IRAF following
procedures described in
the HST Data Handbook.  The wavelength shift in different groups of a
single exposure, as well as between different exposures, was determined with
the task {\bf poffsets} by using
the wavelength information in the associated calibration files
\footnote{The ``standard'' way of using the task {\bf poffsets} misses
out weak and narrow features present in this spectrum. We used it with
"usecorr=no"}.
The groups were then aligned and combined using {\bf specalign}.
Wavelength and flux information was combined using {\bf mkmultispec}.
The resulting total G160M spectrum, smoothed by a boxcar of seven
pixels, is shown in Figure 1.

The G160M spectrum clearly shows multiple components in the absorption
line profile. By using the wavelength difference in the CIV doublet
components, we find that there may be as many as six different
velocity components (Figure 1).
Accurate determination of absorption line parameters is our primary
goal, and so, to determine a reasonable estimate, we used three
different methods to measure the absorption lines. First, we used {\bf
splot} to measure the absorption line equivalent widths (EWs) by
simple linear interpolation across the line in the emission line
profile. This method is unable to measure the blended components 4 and
5 separately.  We next fitted each absorption line profile with a
Gaussian, again in {\bf splot}. These two sets of measurements served
as the initial guesses of the EWs.  To measure the correct EWs, we
have to take into account the emission line profile.

Fitting of the emission and absorption lines together was done using
the STSDAS task {\bf specfit} (Kriss 1994). First we used the lower
resolution G140L spectrum to fit the power-law continuum and the
emission line profile. The emission line was fitted with broad and
narrow Gaussian profiles over the regions unaffected by
absorption. These fit parameters were used in the subsequent fits to
the high resolution spectrum. The parameters of the power-law
continuum, the slope and normalization, were fixed in all the fits to
the high resolution data. The true shape of the emission profile is
the major unknown that affects the EWs of the absorption lines. To
assess the range of probable EWs, we fitted the emission lines with
three different line profiles, `High', `Medium' and `Low', using
slightly different parts of the emission line.
Fits over the entire wavelength range were then performed by adding a
Gaussian for each absorption line. We performed over ten different
fits to obtain estimates of the absorption line parameters. Initially,
component 6 was not included in the fit since this part of the
spectrum might have been a part of the emission line profile. However,
adding this component increased the quality of the fit substantially
with a significant reduction in $\chi^2$. In Table 1, we report the
EWs of the absorption lines for the four most acceptable fits.  In FIT
1, we fixed the parameters of the emission line to the Low profile.
In FIT 2, we fixed these parameters to the Medium profile. In FIT 3,
we allowed these parameters to vary.  The shape of the emission line
profile in this case is close to the High profile (see figure 2).  The
width of both doublet lines in a given component was required to be
the same, although different from component to component; and the
$\Delta\lambda$ was kept constant ( at 2.64 \AA).  Except for these
constraints, the absorption line parameters (EW, centroid, FWHM) were
allowed to be free.  Inspection of these results (Table 1) shows that
the doublet ratios for most of the absorption components are not equal
to the value of two required by atomic physics for optically thin gas
(except for components 3 \& 6 in fit 2). Given the small EW of the
lines, saturation is unlikely to be the cause. For lines with low EWs,
partial covering of the continuum source might be the cause of doublet
ratios close to one. We discuss this in details below.
We also performed one more fit (Fit 4) by fixing the doublet EW ratio
to two, and allowing the emission line parameters to be free to
vary. The shape of the emission line profile in this case is even
higher than the High profile (Fig. 2), which seems unlikely. The fit
in this case is also significantly worse than for fits 1--3.

Fit 2 gives the statistically best fit and in Table 2 we list the EW,
FWHM, centroid wavelength, and the CIV column density inferred for
each component for this case. We note, however, that these values are
not unique since Fit 3 is statistically indistinguishable from Fit 2.
The Fit 2 derived parameters of the emission lines, flux, centroid and
FWHM, are also listed in Table 2.  The large range in the EWs
resulting from different fits (typically a factor 2-3, but even a
factor of 10 in some cases) is clearly seen in Table 1 and indicates
the uncertainty in these measurements. However, as discussed
elsewhere, emission lines profiles in fits 1 and 4 are unlikely to be
true. Considering only fits 2 \& 3, the scatter in EWs range from a
factor of 1 for component 4(1) to a factor of 5 in component 6(2).

\section{The UV Absorbers}

The CIV absorber has six clear, distinct velocity components (marked
as components 1 -- 6 in figure 1). Components 1, 3 \& 6 are broad
(with FWHM$>150$ km s $^{-1}$, Table 2) while components 2, 4, \& 5
are narrow.  Note that the EW of the weak line of component 1 (1(2))
is at \lax 1$\sigma$ level for all fits, except for fit 4 where it was
constrained to be higher. It is therefore possible that component 1 is
spurious. Similarly, EW of component 6(2) is barely at 3$\sigma$ level
for Fit 1. So we believe that Low profile may not be the correct
description of the emission line.  The doublet ratios of the broad
components in Fit 2, the best fit, are close to two, while for the
narrow components the ratio is systematically smaller.  We
investigated whether partial covering might be playing a role for the
narrow components. Following Hamann \etal (1997) we measured the
residual intensity at the troughs of the absorption lines and
determined the covering fraction C$_f$ using their equation 5. We
found C$_f$ to be 0.39, 0.99 \& 0.73 for components 2, 4 \& 5
respectively. Based on flux errors, we estimate errors on C$_f$ to be
7.5\%, 13.5\% and 15\% for components 2, 4 \& 5 respectively. Thus it
is likely that the closely spaced components 4 \& 5 have the same
covering fraction of $\sim 0.85$, with component 2 having about half
that value. `Partial covering' may be due either to a simple
geometrical factor or to `filling in' of the absorption lines by
scattered continuum as in BALQSOs (Cohen \etal~1995).

The CIV column density in each component was calculated by integrating
the optical depths over the absorption line profile (Savage \& Sembach
1991) for (1) complete coverage (C$_f=1$) and (2) partial coverage
(C$_f<1$) assuming constant C$_f$ across the line for narrow lines. We
used equation 6 in Hamann \etal (1997) to calculate optical depths in
case of partial covering. The resulting values of N(CIV) for Fit 2 are
given in Table 2. We also calculated N(CIV) using the measured EWs and
assuming the line to be on the linear part of the curve-of-growth, a
very good assumption given the low EWs and observed FWHM of the lines.
The values of N(CIV) from this method are similar to those obtained
from the optical depth method within 0.1 dex (see Table 2).  The
scatter in N(CIV) from all the fits 1--4 is about 0.2 dex for
component 5 and 0.3 dex for components 2, 3 \& 4. The errors are
larger, about an order of magnitude, for component 6 which is closest
to the emission line peak where the emission line profile is most
uncertain.
Clearly, the uncertainty in column density is governed
by the uncertainty in the underlying emission line profile.

The resolution of GHRS is $\approx$20,000, corresponding to $\sim 15$
km/s at 1575 \AA\. ~So all the absorption lines are resolved with FWHM
ranging from $\sim 40$ km s$^{-1}$ for component 2 to $\sim 300$ km
s$^{-1}$ for component 1.  The thermal width of carbon in a 10$^4$ K
gas is about 4~km/s and of a 10$^5$ K gas is about 12 km/s. Since
photoionization models based on X-ray observations imply temperature
\lax 10$^5$ K, this is about the highest temperature allowed for CIV
absorption systems. The total column densities would otherwise be even
higher.  The implication is that there are nonthermal velocities,
possibly due to bulk or turbulent motions, involved with the absorbing
flows.

Except for components 1 and 6, the absorption lines are deeper than
the power-law continuum emission. The absorbing gas must cover the
BELR, at least partially, and so must lie outside the BELR. The only
other possibility is that the system is absorbing flux from the far
side of the BELR and so may be co-spatial with it.

\section{Discussion \& Conclusions}

The systemic redshift of NGC5548 is uncertain to about hundred km
s$^{-1}$. As reported by De Vaucouleurs \etal (1991), the 21-cm HI
velocity is 5149 km s$^{-1}$ (z=0.017175) while the optical emission
line measurements give 5026 km s$^{-1}$.  Shull \& Sachs (1993) used
5133 km s$^{-1}$ while cz=5220 km s$^{-1}$ was used in the IUE
campaign (Clavel \etal~1991). The observed GHRS redshift of the broad
emission line is cz=4950 km s$^{-1}$ and that of the narrow emission
line is 5130 km s$^{-1}$ (Table 2), consistent with the HI
measurement.  The broad line is blueshifted by 180 km s$^{-1}$ with
respect to the narrow line.

The narrow lines have FWHM\lax 100 km s$^{-1}$ and could arise in the
interstellar medium or halo of the host galaxy.  The broad
components, on the other hand, are likely to be located closer in the
nucleus. The broad components also fully cover the CIV emission
region. A simple geometrical arrangement would be to have the broad
absorbing region outside, but attached, to the BELR and of similar
size.  The redshifts of absorbing components range from cz=3900 km
s$^{-1}$ for component 1 to cz=5400 km s$^{-1}$ for component 6 (Fit
2). While broad components 1 \& 3 indicate material outflowing from
the nucleus with velocities from 1050 \& 450 km/s respectively,
component 6 must be associated with an inflow with a velocity of 450
km/s with respect to the BELR, suggesting an outflow as well as an
inflow in the nuclear region of NGC5548. However, the broad CIV
emission line peak and the absorption systems 1--5 are outflowing
with respect to the host galaxy HI velocity. The component 6 is
inflowing with respect to the HI velocity as well.  There is no
systematic trend of line properties, FWHM or EW, with redshift
offset, possibly due to random motions.  Clearly, the kinematics of
the nuclear region in NGC5548 is complex. Done \& Krolik (1996) have
studied the kinematics of the broad emission line region in NGC 5548
and based on velocity resolved reverberation mapping found evidence
for significant radial infall in addition to random motions. Our
observations of absorption line components support their claim, but
give strong evidence of outflow as well. The infalling component 6
might be related to the accretion flow onto the central black hole.

As discussed earlier, NGC5548 contains an ionized X-ray absorber.  Is
the X-ray absorber related to any of the UV absorbers seen in the GHRS
spectrum?  Based on the model of the X-ray absorber (see MEW95 for
details), the predicted CIV column density is 2--7$\times 10^{13}$
cm$^{-2}$ [Note that this is an order of magnitude smaller than that
derived in MEW95 because of subsequent significant changes in the
photoionization code CLOUDY (Ferland 1996). For the present
calculations we have used the latest version 90.04 of CLOUDY]. Given
the CIV column density in the absorption line components (Table 2),
the X-ray warm absorber is likely to be associated with one of the
broad components, either 3 or 6. [However, this conclusion could
change if further updates to CLOUDY produced factor of 3 change in
predicted N(CIV)]. If component 6 is indeed related to the accretion
flow and thus is cold, then the X-ray ionized absorber may likely be
related to component 3.

The definite way to resolve this is with high resolution X-ray
spectroscopy.  As yet there is no X-ray spectroscopy anywhere close to
GHRS resolution, so direct tests are not possible. The broad
components 3 \& 6 are separated by $\Delta z= 0.003$ corresponding to
$\sim 2$ eV at the OVII edge of the warm absorber. High resolution
observations with AXAF would be able to determine the edge energy
accurately enough to tell them apart. The absence of greater X-ray
absorption implies that the remaining CIV components must have a much
lower ionization state and smaller total column density by at least an
order of magnitude.

Kinematically complex UV absorption, similar to that reported here, is
also present in other Seyfert galaxies, e.g. Mrk 509 (Crenshaw \etal
{}~1995), NGC4151 (Weymann \etal ~1997), NGC3516 (Kriss \etal 1996,
Crenshaw \etal 1999). All of these objects also show presence of a
X-ray warm absorber, qualitatively similar to that in NGC5548
(Reynolds 1997, Warwick \etal ~1996, Mathur \etal 1997 for Mrk509,
NGC4151 and NGC3516 respectively). In Mrk509 and NGC4151, parameters
of the warm absorber are not well constrained due to the complexity in
X-ray spectra; as a result comparison with UV absorption is difficult.
The UV absorption system in NGC3516 also shows narrow as well as broad
components. However, the broad components have disappeared since
$\sim1992$. Mathur \etal~(1997) have shown that an evolving X-ray/UV
absorber is consistent with the presence of X-ray warm absorber,
presence of broad UV absorption in the IUE data and their
disappearance in the recent UV observations. Our GHRS observations of
NGC5548 thus marks the first case in which complex UV absorption is
present and for which a meaningful comparison with X-ray absorption
can be made. It should also be noted that the UV absorption in a high
resolution GHRS observation of NGC3783 is also found to be consistent
with X-ray absorption (Shields and Hamann 1997), though the UV
absorption in that case is not kinematically complex.

Even though our unified X/UV absorption scenario (MEW95 and references
therein) has stood the test of high resolution observations, it may
not be adequate to explain complex systems in its present single-zone,
photoionization equlibrium form. Multi-zone, multiparameter models
will be required when the absorption systems show multiple components
(e.g. objects in Hamann 1997). In some highly variable objects
(e.g. NGC4051), non-equilibrium models might be necessary (Nicastro
\etal 1999). The X/UV models should evolve and become more refined as
better quality data become available (see Mathur 1997). At present,
all the available data (Mathur \etal~1998) clearly suggests that the
UV and X-ray absorbers are physically related and perhaps identical;
at a minimum the X-ray absorption makes a substantial contribution to
the absorption seen in the UV.

We greatly appreciate the help of Michele De La Pena, Claus Leitherer
\& Stefi Baum of STScI with the GHRS data analysis. We thank Mike
Crenshaw for useful discussions.  SM acknowledges financial support
through NASA grants NAG5-3249 (LTSA) and GO-06485.01-95A from Space
Telescope Science Institute. BJW is supported through NASA contract
NAS8-39073 (ASC).


\newpage
\noindent
{\bf Figure Captions:}\\

\noindent
{\bf Figure 1:} The GHRS G160M spectrum of CIV line in NGC5548. It is
smoothed to display
the multiple absorption components clearly (marked 1--6).

\noindent
{\bf Figure 2:} Four different line profiles fitted to and superposed
upon the CIV
emission line.

\noindent
{\bf Figure 3:} Superposition of the four fits to the emission $+$
absorption lines on the data.

\newpage
\thispagestyle{empty}

\begin{table}[h]
\caption{~Observed Absorption Line EWs in \AA.~}
\begin{tabular}{|lcccccccccccc|}
\hline\hline
Components$^a$& 1(1) & 1(2) & 2(1) & 2(2) & 3(1) & 3(2) & 4(1) & 4(2) & 5(1) &
5(2)
& 6(1) & 6(2) \\
\hline
Fit 1 & 0.1& 0.003& 0.08& 0.06& 0.3& 0.1& 0.38& 0.28& 0.29&
0.22& 0.01& 0.04 \\
Fit 2 & 0.11& 0.005& 0.08& 0.07& 0.33& 0.17& 0.33& 0.24& 0.38& 0.30&
0.13& 0.06 \\
Fit 3 & 0.09& 0.01& 0.09& 0.08& 0.44& 0.25& 0.33& 0.25& 0.41& 0.29&
0.45& 0.37 \\
Fit 4 & 0.07& 0.035& 0.09& 0.04& 0.2& 0.1& 0.30& 0.15& 0.43& 0.21&
0.82& 0.41 \\
\hline
\hline
\end{tabular}
\smallskip
\small
a. (1) and (2) represent shorter and longer wavelength components of
 a doublet.\\
\normalsize
\end{table}
\newpage
\thispagestyle{empty}

\vspace*{1.0in}
\begin{table}[h]
\vspace*{6.1in}
\begin{rotate}{90}
{\bf Table 2:}~Absorption and Emission Line Parameters for Fit 2
(Medium emission line profile).
\end{rotate}
\hspace*{2.5in}
\begin{rotate}{90}
\begin{tabular}{|lcccccccccccc|}
\hline\hline
Absorption& 1(1) & 1(2) & 2(1) & 2(2) & 3(1) & 3(2) & 4(1) & 4(2) & 5(1) & 5(2)
& 6(1) & 6(2) \\
Components&&&&&&&&&&&& \\
\hline
EW & 0.11& 0.005& 0.08& 0.07& 0.33& 0.17& 0.33& 0.24& 0.38& 0.30&
0.13& 0.06 \\
 FWHM$^a$& 291& 291& 39& 39& 165& 165& 67& 67& 103& 103& 168& 168 \\
 Centroid& 68.70& 71.32& 70.82& 73.43& 71.54&
74.16& 72.39& 75.01& 72.75& 75.36& 76.09& 78.70 \\
$\lambda$1500$+^f$&&&&&&&&&&&& \\
cz$^{af}$ & 3900& & 4380& & 4500& & 4680& & 4770& & 5400& \\
$log$N$^b$&& 12.4&& 13.5&& 13.9&& 14.1&& 14.2&& 13.6 \\
$log$N$^c$(C$_f=1$)&& 12.4&& 13.6&& 13.9&& 14.2&& 14.3&& 13.4 \\
$log$N$^d$(C$_f<1$)&& && 14.2&& && 14.2&& 14.4&&  \\
&&&&&&&&&&&& \\
\hline
Emission Line & \multicolumn{3}{c}{Flux$^e$} &
\multicolumn{3}{c}{Centroid$^f$} & \multicolumn{3}{c}{cz$^{af}$} &
\multicolumn{3}{c|}{FWHM$^a$} \\
Broad & \multicolumn{3}{c}{5.9E-12} & \multicolumn{3}{c}{1574.73} &
\multicolumn{3}{c}{4950} & \multicolumn{3}{c|}{6978.85}\\
Narrow & \multicolumn{3}{c}{8.6E-13} & \multicolumn{3}{c}{1575.60} &
\multicolumn{3}{c}{5130} & \multicolumn{3}{c|}{1074.18} \\
\hline
\hline
\end{tabular}
\end{rotate}
\small
\hspace*{2.3in}
\begin{rotate}{90}
a: in units of km/s.\\
\end{rotate}
\hspace*{0.1in}
\begin{rotate}{90}
b: from EWs, assuming linear part of curve-of-growth.  \\
\end{rotate}
\hspace*{0.1in}
\begin{rotate}{90}
c: from integrating optical depths over the line profile, assuming
covering fraction of 1.\\
\end{rotate}
\hspace*{0.1in}
\begin{rotate}{90}
d: using a constant covering fraction $<1$ across the line for narrow
lines. See text for values of C$_f$\\
\end{rotate}
\hspace*{0.1in}
\begin{rotate}{90}
e: in units of ergs s$^{-1}$ cm$^{-2}$ \AA$^{-1}$\\
\end{rotate}
\hspace*{0.1in}
\begin{rotate}{90}
f: these are as measured form the {\bf specfit} fit.  We estimate
 systematic wavelength offset of $\sim0.095$\AA\ \\
\end{rotate}
\hspace*{0.1in}
\begin{rotate}{90}
 or a redshift  by $\Delta$cz$\sim18$ km s$^{-1}$ caused by the
 manipulation to the raw data.\\
\end{rotate}
\end{table}
\normalsize

\newpage
\begin{figure}
\psfig{figure=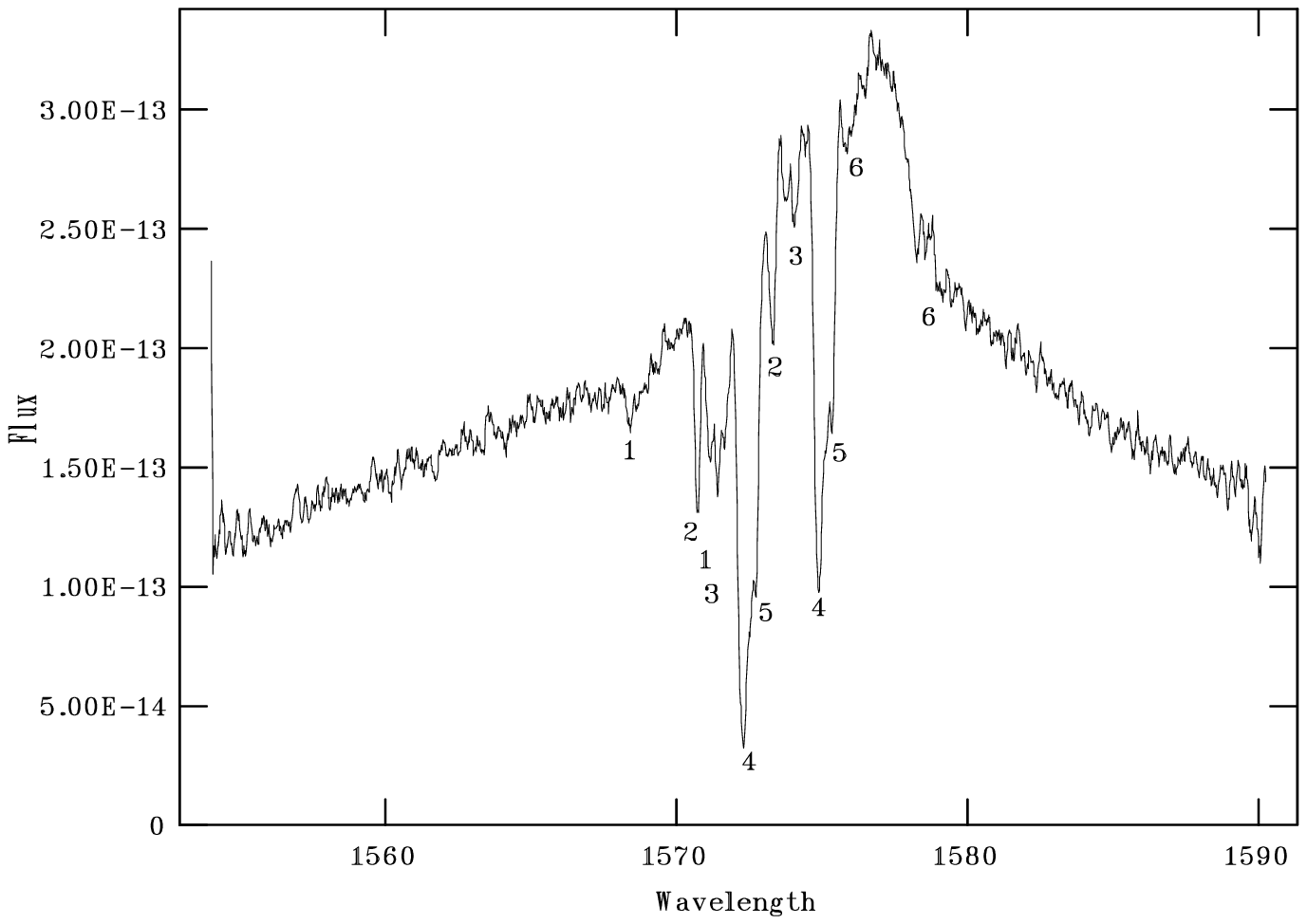}
\end{figure}

\newpage
\begin{figure}
\psfig{figure=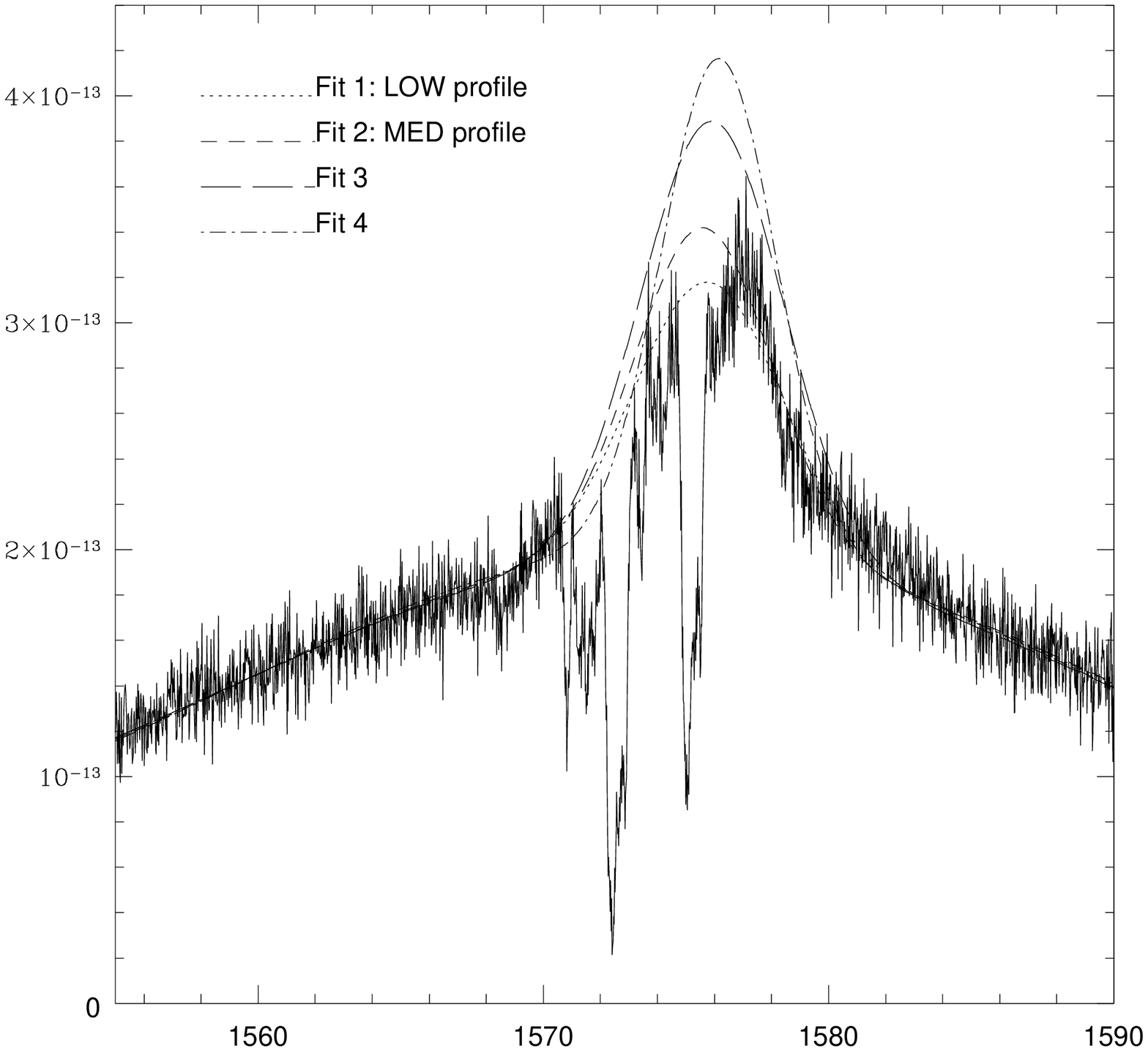,height=5truein}
\end{figure}

\newpage
\begin{figure}
\psfig{figure=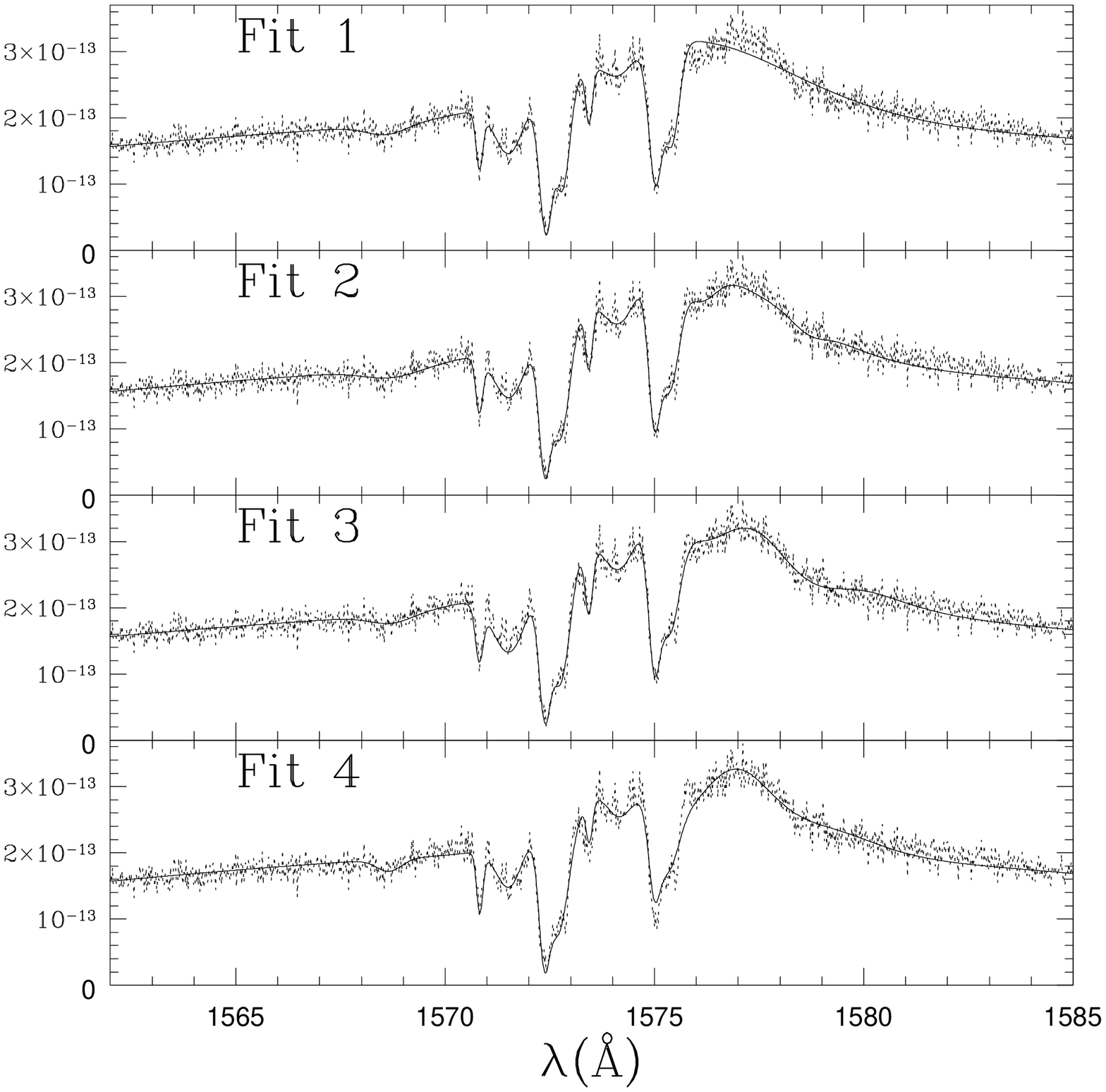,height=5truein,width=4in}
\end{figure}

\end{document}